\documentclass[12pt,preprint]{aastex}

\usepackage{rotating}

\def\mum ~{{$\rm \mu$m}}

\shorttitle{Gaps in the cloud cover.}
\shortauthors{Holwerda et al.}

\begin{document}

\title{Gaps in the cloud cover?\\
Comparing extinction measures in spiral disks}
\author{B. W. Holwerda\altaffilmark{1}, M. Meyer\altaffilmark{1}, M. Regan\altaffilmark{1},  D. Calzetti\altaffilmark{2}, K. D. Gordon\altaffilmark{3}, J. D. Smith\altaffilmark{3}, D. Dale\altaffilmark{4}, C. W. Engelbracht\altaffilmark{3}, T. Jarrett\altaffilmark{5}, M. Thornley\altaffilmark{6}, C. Bot\altaffilmark{7}, B. Buckalew\altaffilmark{8}, R. C. Kennicutt\altaffilmark{9}, R. A. Gonz\'{a}lez \altaffilmark{10}}

\altaffiltext{1}{Space Telescope Science Institute, 3700 San Martin Drive, Baltimore, MD 21218, USA}
\altaffiltext{2}{University of Massachusetts, Department of Astronomy, 710 North Pleasant Street, Amherst, MA 01003, USA}
\altaffiltext{3}{Steward Observatory, University of Arizona, 933 N Cherry Ave, Tucson, AZ 85721, USA}
\altaffiltext{4}{Dept. of Physics \& Astronomy, University of Wyoming, Laramie, WY 82071}
\altaffiltext{5}{Spitzer Science Center, MS 220-6, Caltech, 1200 E California blvd. Pasadena CA 91125, USA}
\altaffiltext{6}{Bucknell University, Lewisburg, PA 17837, USA}
\altaffiltext{7}{California Institute of Technology, MC-320-47, 1200 E California blvd. Pasadena CA, USA }\altaffiltext{8}{Dept. of Physics, Embry-Riddle University, 3700 Willow Creek Rd, Prescott, AZ  86301, USA}
\altaffiltext{9}{Institute of Astronomy, University of Cambridge, Madingley Road, Cambridge CB3 OHA, UK}
\altaffiltext{10}{Centro de Radiastronom\'{\i}a y Astrof\'{\i}sica, Universidad Nacional Aut\'{o}noma de M\'{e}xico, 58190 Morelia, Michoac\'{a}n, Mexico}

\email{holwerda@stsci.edu}

\begin{abstract}
Dust in galaxies can be mapped by either the FIR/sub-mm emission, 
the optical or infrared reddening of starlight, or the extinction of a known 
background source.
We compare two dust extinction measurements for a set of fifteen sections in 
thirteen nearby galaxies, to determine the scale of the dusty ISM 
responsible for disk opacity: one using stellar reddening and the other a known background 
source.
In our earlier papers, we presented extinction measurements of 29 galaxies, 
based on calibrated counts of distant background objects identified though 
foreground disks in HST/WFPC2 images. 
For the 13 galaxies that overlap with the {\it Spitzer} Infrared Nearby 
Galaxies Survey (SINGS), we now compare these results with those obtained 
from an $I-L$ color map.
Our goal is to determine whether or not a detected distant galaxy indicates 
a gap in the dusty ISM, and hence to better understand the nature and geometry of 
the disk extinction. 

We find that distant galaxies are predominantly in low-extinction 
sections marked by the color maps, indicating that their number depends 
both on the cloud cover of {\it Spitzer}-resolved dust structures --mostly the spiral arms--
and a diffuse, unresolved underlying disk. We note that our infrared color map (E[I-L]) 
underestimates the overall dust presence in these disks severely, because it 
implicitly assumes the presence of a dust screen in front of the stellar distribution.
\end{abstract}

	
\section{\label{sec:intro}Introduction}

The dust content of a spiral galaxy can be estimated using several different 
observational techniques: 
(1) observations of the dust's typical emission in the far-infrared and sub-mm 
spectral ranges \citep[e.g.,][]{Dale05,Meijerink05};
(2) the reddening and dimming effect on stellar populations, in the UV \citep[e.g.,][]{Boissier04}, 
optical \citep[e.g.,][]{Elmegreen80} and infrared \citep[e.g.,][]{Regan00,Martin06};
(3) the balancing of the disk's energy budget that produces its spectral energy 
distribution \citep[e.g.,][]{Cortese05,Popescu05rev};
or (4) the absolute absorption of a {\it known} background source 
\citep[e.g.,][]{kw99a,Gonzalez98,Elmegreen01, Holwerda05}. 
Reviews of the various approaches can be found in \cite{Calzetti01, Popescu05rev, Tuffs05rev} 
and references therein.

All of these techniques have different and specific strengths and weaknesses and 
hence a comparison between the results of two such techniques offer insight into 
the nature of the dust in disks. In two recent papers \citep{Holwerda06sings,Holwerda07a}, 
we compare the dust surface densities estimated from distant galaxy counts and the 
star/dust spectral energy distribution (SED). Here, we focus on the comparison 
between extinction maps based on stellar reddening, and results from the distant 
background object counts. The reddening map's advantage over the SED 
measurement is spatial resolution. 
The simple screen geometry we use here has been surpassed in extinction studies 
for some time \citep[e.g.,][]{Elmegreen80, Kodaira94}, but will suffice to determine 
the scale of the dusty ISM structure that is responsible reducing the number of distant 
galaxies seen through a spiral disk.
Our goal is to determine whether an observed distant galaxy indicates a hole in the 
ISM disk that would correspond to no extinction in the reddening map.

An independent extinction measurement requires a known background 
source other than the stellar disk. \cite{kw92} proposed an occulted 
elliptical galaxy, and \cite{Gonzalez98} used the calibrated number of distant 
galaxies seen through a foreground disk in a Hubble Space Telescope image. 
Both techniques have been used on nearby disks, and have by now exhausted 
nearby occulting pairs \citep{Andredakis92,Berlind97,kw99a,kw00a,kw00b,kw01a,kw01b,Elmegreen01}
and HST/WFPC2 images suitable for distant galaxy counts \citep{Holwerda05a, 
Holwerda05b, Holwerda05c, Holwerda05e, Holwerda05d}. 
The known background source is either rare (the occulted large galaxy) or 
inherently uncertain due to cosmic variance (number density of distant background 
galaxies). The advantages are that no assumption about the dust/star 
geometry has to be made, and the absorption measurement is for the entire height 
of the disk.

Models for the light of spiral disks have incorporated ever more complex geometries 
for extinction and reddening by dust, both for edge-on systems \citep[e.g.,][]{Kodaira94, Xilouris99, Kylafis05} 
and face-on spirals \citep[e.g.,][]{Elmegreen80, Elmegreen99, Disney89, Rix93, Peletier95}. 
These have been used to correct stellar populations \citep{Elmegreen80, Rix93} and star 
formation \citep[See the review of ][and references therein]{Calzetti01}. The large scale 
structure of the dusty ISM in spirals has been modeled this way with a dusty disk more 
extended but thinner than the stellar one \citep{Elmegreen80, Peletier95, Xilouris99, Bianchi07}. 
\cite{Dalcanton04} argue that dust lanes are a sign of vertical instability 
in more massive disks while smaller disks exhibit a fractured dust morphology. Recent 
observations point to a thicker or second vertical structure of dust ISM \citep{Howk99a,Howk99b,
Thompson04,Seth05, Kamphuis07}.

Monte-Carlo simulations of photons in the disk also use ever more sophisticated geometry: 
spherical \citep{Witt92, Witt00}, cylindrical \citep{Bianchi96}, planar \citep{Bruzual88,Baes01a, 
Baes01b}, and recently arbitrary geometry \citep[e.g.,][]{Gordon01, Misselt01, Baes03}. 
The clumpiness of the ISM has been proven to be the chief reason why the observed extinction 
law differs from the Galactic one \citep{Natta84,Calzetti94,Fischera03,Fischera05}.
The aim of models has shifted to predicting not only the reddening and extinction by the dust 
in spirals, but also the contribution from the ISM's emission to the overall spectral energy distribution. 

Multi-wavelength models and observations of disks are addressing several questions 
regarding spiral disks: star-formation measurements \citep[e.g.,][]{Calzetti05, Dopita06a,Dopita06b}, ISM 
composition \citep[e.g.,][]{Draine01a,Draine01b,Draine06,Draine07}, and energy balance 
\citep{Popescu00,Misiriotis01,Tuffs04,Dasyra05}, and stellar populations \citep{Dale05,Dale07}.

The results from the {\it Spitzer} Infrared Nearby Galaxy Survey \citep[SINGS,][]{sings}
have shed light on the relation between dust and star-formation within a range of galaxy types 
and environments: the starburst M51 \citep[]{Calzetti05,Thornley06}, the grand design spiral 
M81 \citep{Gordon04, PG06}, the rings of NGC 7331 \citep{Regan04} and M31 \citep{Gordon06}, 
the superwind in M82 \citep{Engelbracht06}, and the dwarf NGC 55 \citep{Engelbracht04}. 
Comparative studies for the whole sample of galaxies can be found in \cite{Dale05,Dale07}, 
and \cite{Draine07}.

The mid-infrared bands at 3.6 \mum ~  and 4.5 \mum ~ are dominated by 
the emission from the older, smooth stellar disk; either combined with 
the $I$-band data they can provide a very useful dust screen extinction map
\citep[][]{Regan00,Martin06}. We employ the simple screen approach 
because (1) in the near and mid-infrared, we expect most of the disk to 
be optically thin, with the Spitzer bands providing the lowest extinction view 
of the disk, and (2) we are only interested in relative opacity values. 

The advantage of the color-based extinction map is that the spatial resolution is 
as high as the original single band images; however the derived extinction values are 
limited by the assumptions about geometry (a single screen of dust in 
front of a layer of stars) and stellar population (single stellar color everywhere).

We want to know whether or not a disk is fully transparent in those places where a distant galaxy 
can be identified in HST images. The notion that those lines-of-sight are devoid of dust seems 
to be in contradiction with the picture of an ubiquitous, fractal and clumpy medium has developed since 
the discussions on the opacity of spiral disks \citep{Cardith94}.

The organization of this paper is as follows: \S \ref{sec:data} describes 
the data, \S \ref{sec:ext} details the calibrations of the extinction maps, \S 
\ref{sec:sfm} briefly describes the distant galaxy counts, and \S 
\ref{sec:comp} and \S \ref{sec:concl} are our discussion and conclusions.

\section{\label{sec:data}Data}

In this paper we use the {\it Spitzer}/IRAC mosaics, and accompanying 
ground-based $I$-band data, from the fourth data-release of the SINGS project 
\citep{sings,dr4}. Part of the {\it Spitzer} data products are mosaics of the 4 
IRAC channels with a pixel scale of $0\farcs75$ in $\rm MJy ~ sr^{-1}$, aligned 
with the sky coordinates. The mosaics are the product of a specialized pipeline to 
register and ``drizzle'' the basic calibrated data to a single mosaic.

The ground-based $I$-band images are a complementary SINGS data product. 
The world coordinates of the $I$-band data are updated using the stellar positions 
in the IRAC, channel 1 (3.6 \mum ~) mosaic. The $I$-band images are then shifted to the 
IRAC mosaic coordinates and degraded to the same resolution --estimated from the stars 
in the field, generally with a FWHM of $2\farcs5$. 

\section{\label{sec:ext}Extinction maps}


The {\it Spitzer}/IRAC channels 1 and 2 (3.6 and 4.5 \mum ~) are mostly devoid of the spectral 
features of interstellar Polycyclic Aromatic Hydrocarbons (PAH) and suffer little from extinction 
by the larger dust grains. In this paper we use channel 1 (3.6 \mum ~) to map the stellar emission.
The I-L color is proportional to the extinction in $I$ ($A_I$):

\begin{equation}
\label{eq:extlaw}
A_I =  a \times (E[I-L]  + b).
\end{equation}

\noindent This leaves two free parameters for the NIR extinction law: the zero-point, $b$,
and the slope, $a$. The slope of the IR extinction law is very close to unity: $a = 0.88$ 
\citep{Rieke85} for single lines-of-sight. The zero-point $b$ can be found using the 
IRAC 8 micron PAH-map of the same galaxy\footnote{The 8 micron PAH-only map is 
obtained from IRAC's channel 4, by subtracting the stellar contribution to it. This contribution 
is estimated from channels 1 and 2 (3.6 and 4.5 \mum ~), using the same formalism as 
\cite{Pahre04a,Pahre04b} but with the improved aperture corrections from 
\cite{IRACaper} and \cite{Dale07}.} 

To determine the zero-point ($b$) we find the 100 lowest-value pixels in the 8 \mum ~ 
PAH-map, after sigma-clipping to remove the sky. The median value of the 
100 corresponding pixels in the extinction map is our zero-point $b$, ie., the stellar color 
of the disk before extinction. Table \ref{tab:b} lists the zero-extinction colors for our maps.

The extinction maps are clipped to the 1-sigma level calculated from the sky noise in both 
$I$ and $L$ images. The spatial resolution of the extinction maps is predominantly limited 
by the $I$-band observations; FWHM estimates of the PSF are around $2\farcs5$.


In the construction of the extinction maps, we make three significant assumptions: 
(1) the geometry of the dust and stars is a simple screen of dust in front of the stars; 
(2) regions lacking 8 \mum ~ emission are indicative of the intrinsic stellar color; and 
(3) the $L$-band (3.6 \mum ~) is a good tracer of purely stellar emission. As noted 
in the introduction, more sophisticated models exist for the dust's effect on stellar emission 
but these maps will suffice for our purposes. Similar maps are used by \cite{Martin06} to 
trace the dusty ISM for comparison with the morphology of CO and HI maps.

The first assumption is a screen of dust in front of the stars. This is the simplest geometry 
one could use, and it allows the application of the extinction law as found for single lines-of-sight. 
Mixing the extincting dust and stars would result in a decoupling of the relation between color 
and extinction  \citep{Natta84,Calzetti94,Fischera03,Fischera05}.
%
Alternative maps could be constructed using a more gray relation than equation \ref{eq:extlaw}, 
reflecting the mix of dust and stars. Unfortunately, the choice of extinction law is problematic, as 
the exact mix and geometry are unknown and likely to change in different sections of the disk. 
However, we are interested in the relative extinction in our maps (whether or not distant galaxies 
are preferentially detected in low-extinction regions) and not in the absolute extinction along a given 
line-of-sight. We note that these maps represent lower limits for the total extinction through the disk.

The second assumption is that PAH emission does not arise in sections that have no 
interstellar dust; therefore, these sections will exhibit no reddening in the E[I-L] map.
Recent observations with {\it Spitzer} have found regions in galaxies with 24 \mum ~  emission 
but no 8 \mum ~  emission \citep[e.g.,][]{sage}. This is attributed to different processing of the dust 
grains and PAHs by the strong UV fields of star-forming regions. However, the lowest fluxes of 
PAH emission in these images occur not in the centers of bright star-forming regions but in the disk, 
between the arms. 

The third assumption is the choice of reference band for the purely stellar emission.
\cite{Martin06} chose channel 2 (4.5 \mum ~) as the NIR reference, and in this paper we 
use channel 1 (3.6 \mum ~). Both suffer small scale contamination from non-stellar 
emission; channel 2 (4.5 \mum ~) may contain hot dust emission \citep{Lu03,Pahre04a,
Regan04,Calzetti05,Hunter06}, and channel 1 (3.6 \mum ~) comprises a smaller PAH 
emission feature, centered near 3.3 \mum ~, and may contain hot dust emission as well.
In both cases, the PAH or hot dust would increase the flux in the ``stellar'' channel, 
artificially elevating the implied extinction in the $I$-band. This additional uncertainty 
is of the order of the stellar color gradient (approximately a tenth of a magnitude).
Both uncertainties --contamination and stellar gradient-- are much smaller than the 
uncertainty due to the assumed geometry of the stars and dust in the extinction values 
in the presented maps but they do change the distribution. 
To check for systematic effects between channel 1 and 2, we compared our extinction 
map of NGC 3627 made from channel 1 (3.6 \mum ~, Figure \ref{fig:extmaps}) to one 
using channel 2 (4.5 \mum ~) from \citep[][, their Figure 3]{Martin06}).
The comparison reveals little difference in structure --some disparities in the 
HII regions in the spiral arms-- and an overall offset of 0.05 mag in extinction. 
Because the PAHs affect only HII regions and we are interested in the overall disk, 
this should not strongly  influence our overall conclusions.

Given the above assumptions --the star/dust geometry, the intrinsic stellar color of the disk, 
and the link between lack of extinction and absence of 8 \mum ~ PAH emission--, the absolute 
calibration of the extinction maps is uncertain. The dominant presupposition is that of the 
geometry (and hence the extinction law), which most certainly results in an underestimate of 
the dust in these galaxies. 
%
However, because we investigate the ``preference'' of distant objects for low or high extinction 
regions, these maps will suffice to distinguish between sections of higher and lower extinction. 

\section{\label{sec:sfm}Distant galaxy counts}

\cite{Holwerda05b} present disk extinction measurements based on counts of distant galaxies, 
calibrated for crowding and confusion with the ``Synthetic Field Method'' \citep[SFM, ][]{Gonzalez98, 
Holwerda05a}. The calibration is done with a series of` ``synthetic fields", the original image 
with an artificially dimmed Hubble Deep Field (HDF) added to it. The relation between artificial 
dimming and HDF galaxies retrieved from the synthetic field can then be used to find the average 
extinction in the field from the number of {\it observed} distant galaxies found in the original field.
The relation between the number of HDF galaxies and dimming is unique for each field, and 
synthetic fields are made for each science field. In the zero-extinction synthetic field, the added 
HDF galaxies show where distant galaxies {\it could} have been detected under the crowding 
conditions in the science field. The SFM gives an average opacity for the whole height of a section 
of the spiral disk \footnote{The term disk opacity is defined as the apparent optical depth for the whole 
height of the disk, expressed in magnitudes.}.

In \cite{Holwerda05b}, we already speculated that if the galaxies identified in the WFPC2 field are in 
those sections of the disk that are effectively transparent, then the average extinction found from 
the SFM is an indication of dust cloud cover. If so, we found that typical covering fractions are 40\% 
for the disk and 60\% for the spiral arms, assuming fully opaque clouds. 
We have compared these disk opacity measurements with those from the occulting galaxy technique 
\citep{Holwerda05b}, and those provided by an SED model based on the IRAC and MIPS data 
\citep{Holwerda07a}. The disk opacity values --apparent optical depths-- for all three methods agree well.
In \cite{Holwerda07a}, we found a typical cloud optical depth of 0.4 that corresponds to a typical cloud size of ~60 pc. 

In this paper, we explore the earlier assertion that distant galaxies are predominantly found in those 
sections of the disk that are completely transparent, something that appears somewhat contradictionary 
to the small scale found in our comparison with the SED. Therefore, we will quantify to what extent 
large dust clouds --resolved with IRAC-- determine the SFM opacity value.

\section{\label{sec:comp}Distant galaxy number and disk extinction}

\begin{figure}
\centering
\plotone{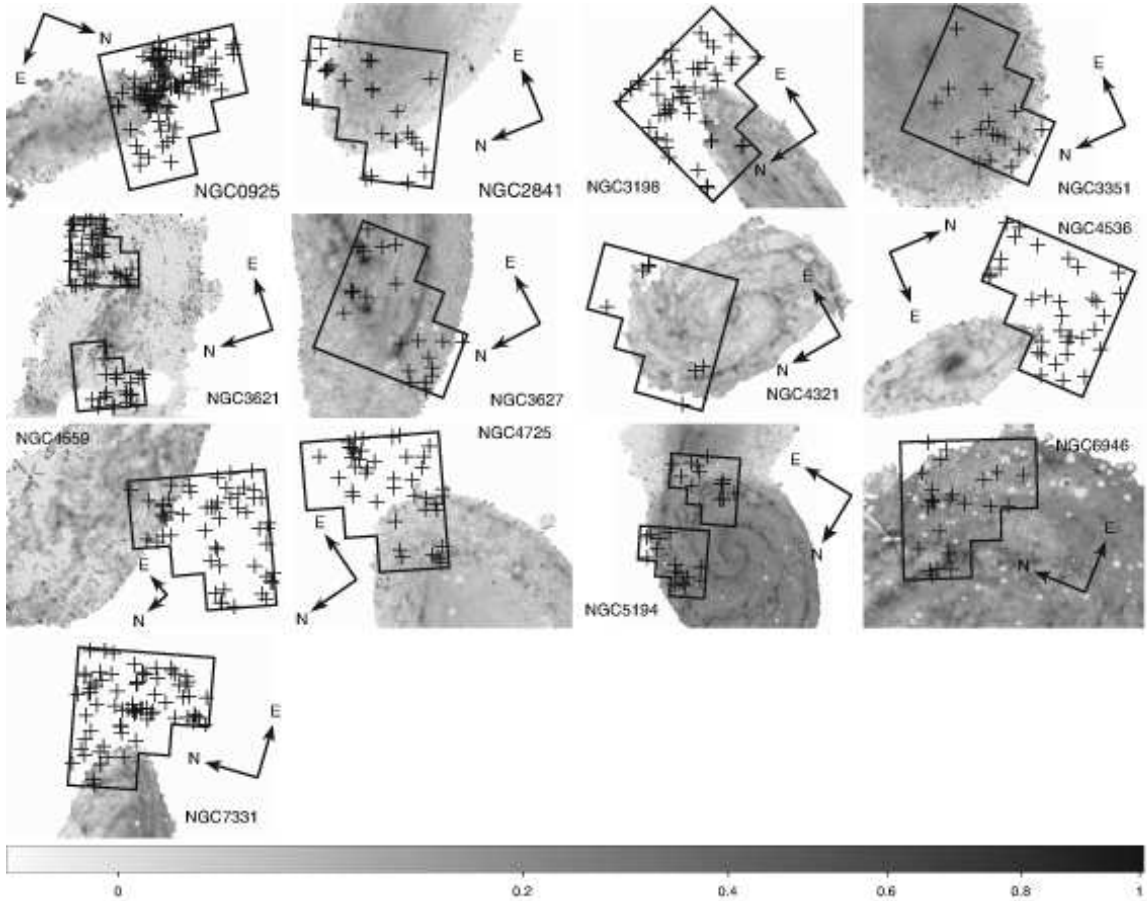}
\caption{\label{fig:extmaps}Details from the extinction maps, based on the E[I-L] color, for the 13 SINGS galaxies in our sample. Grayscale values are the optical depth $\tau$ implied by the I-L color. The WFPC2 field-of-view used for the galaxy counts in \cite{Holwerda05b} is overlaid. The crosses are the positions of the distant galaxies identified in the HST/WFPC2 data.}
\end{figure}

The $I$-band extinction maps based on the above method are presented in Figure \ref{fig:extmaps}.
The extinction is higher in the spiral arms, and the effects of HII regions can clearly be seen. 
In Figure \ref{fig:extmaps}, the WFPC2 footprint and the positions of the identified distant galaxies 
from \cite{Holwerda05b} are also plotted on the extinction map. Since the WFPC2 fields 
are centered on a spiral arm further out in the disk, the sigma-clipped extinction maps often 
do not cover the entire field-of-view of the WFPC2.\footnote{In the case of NGC 3198 and 
NGC 4536, the overlap of the WFPC2 field and the extinction map is small (Figure \ref{fig:extmaps}), 
and these two galaxies do not contribute much to the statistical result below (Figure \ref{fig:hist}).} 

The distant galaxies identified by \cite{Holwerda05b} appear to be mostly in the 
regions of lower extinction (Figure \ref{fig:extmaps}). 
To test this, we compare the distribution of extinction values at the positions of observed 
distant background galaxies to the distribution of extinction values where a distant 
galaxy {\it could} be found, i.e., the position of the synthetic galaxies. 
For the reference distribution, we use the extinction values at the positions of the added 
HDF distant galaxies in the zero-extinction synthetic field. These are positions in the 
extinction map where we {\it could} have found a distant galaxy, given the crowding 
and confusion conditions in the HST/WFPC2 images.

Figure \ref{fig:hist} shows the cumulative histogram for these two sets of extinction 
values, at the position of the observed distant galaxies and the synthetic ones. 
The majority of the observed distant galaxies is detected predominantly below extinction values 
of 0.2 magnitude. Artificial, synthetic distant galaxies can be found in regions with larger 
extinction values. A Kolmogorov-Smirnov test shows that the probability that the two 
distributions are similar is negligible ($\rm 1.98~ 10^{-14}$). Hence, an observed distant 
galaxy is more likely to be found in a low-extinction section of the disk\footnote{The limiting factor 
in the detection of HDF galaxies is the ``granularity'' in the science field, i.e. how resolved the stellar 
disk is \citep[See our discussion in][]{Gonzalez03,Holwerda05d}, not the disk's opacity.  }

Cloud cover by optically thick clouds, which can be resolved with {\it Spitzer}, is one factor reducing 
the detected number of distant galaxies observed through a disk. However, distant background 
galaxies are not found exclusively at zero-extinction regions; 60\% of the observed galaxies are 
identified at more than 0.1 mag of extinction (see Figure \ref{fig:hist}). 
Therefore, a second factor determining their observed number is the lower-extinction, unresolved 
dust disk. If this is a cloud filling factor as well (discrete clouds rather than a screen), these are many 
smaller clouds in addition to the structures seen in the extinction maps, consistent with our result in \cite{Holwerda07a}.

How much of the opacity measured from missing distant galaxies is caused by complete 
blocking of the line-of-sight depends on what value in the extinction map from the infrared 
color translates into a distant galaxy thats is unidentifiable as such.
The extinction maps in Figure \ref{fig:extmaps} underestimate the overall dust surface 
density and hence opacity of the disks. This can be illustrated with a comparison between 
the average opacities (apparent optical depths) from the E[I-L] extinction maps ($A_{[I-L]}$), 
the overall Spitzer SED ($A_{SED}$), and the number of distant galaxies ($A_{SFM}$). 
The average extinction values from the maps --from the part where there is data-- are 
generally much lower than those derived from the SED or the number of distant galaxies 
(Table \ref{tab:ext}).
This can partially be explained by the fact that the SFM measures the extinction for the 
entire height of the disk --the background sources are well beyond the disk-- and the 
color map measures the extinction by at best half of of the disk's height --the part backlit 
by the stellar disk. This would explain a factor of two difference between the two opacity values.
The second reason that the extinction maps underestimate the total opacity is the assumed 
star/dust geometry. A mix of stars and dust does not follow a neat extinction law and will 
generally display a variety of extinction law behaviors, all of them more ``gray'' than the 
one from \cite{Rieke85}, depending on geometry \citep[As shown previously by several 
authors, notably][]{Elmegreen80,Natta84}. Let us set values in the E[I-L] map 
that would be opaque for a distant galaxy and compare how the cloud cover explains 
the missing distant galaxies.

In our E[I-L] extinction maps, $A_{[I-L]}=0.5$ (half the height of the disk), possibly $A_{[I-L]}=0.3$ 
(a third of the height) is the naively expected opaque disk value. Table \ref{tab:f} 
lists the covering percentage of the extinction maps for pixels above $A_{[I-L]} = $0.5, 0.3, and 0.1 
for the whole map and the section covered by the WFPC2. 
Extinction values greater than either 0.5 and 0.3 fail to cover the fraction implied by the 
SFM measurement. Only if the $A_{[I-L]}$ extinction is underestimated by a factor of 10, do the cloud 
cover fractions match up to the same order, yet we found most of our {\it real} observed distant 
galaxies at positions with 0.1 mag of extinction (Figure \ref{fig:hist}). 
Hence, if we take $A_{[I-L]} \approx 0.3$ as the point where the disk is opaque for distant 
galaxies, the big opaque clouds only block $\sim$10\% of the typical background of 
distant galaxies outright (the solid line in Figure \ref{fig:hist}). That corresponds to $\sim$0.3 
mag of the SFM opacity measurement. Hence the remainder --often the majority-- of the 
measured opacity is due to the unresolved and semi-transparent disk.

In \cite{Holwerda05b}, we voiced our suspicion that the distant galaxies are detected 
in gaps in the ISM disk, based on the average color of these galaxies, which 
remained constant with the disk opacity inferred from their number.
In addition, we found that the opacity from the counts of distant galaxies
depended little on the inclination of the foreground disk.

This result we initially explained by a flat distribution of clouds for which the 
projected filling factor remains constant. This model did not specify the typical size of 
the opaque clouds, but a similar to or greater angular size than the projected distant 
galaxies was implied.
In \cite{Holwerda07a}, we found small, optically thin clouds dominate the disks, 
using extra information from the Spitzer SED. Figure \ref{fig:hist} appears to corroborate 
the hypothesis that the filling factor of large, resolved clouds do play a role in determining 
the number of distant galaxies found, but they are not the sole cause for their diminishing number 
and hence extinction in the disk. The second factor is the underlying disk of unresolved, small clouds.
Therefore, the presence of a distant galaxy seen through a foreground disk does not imply 
that no dusty ISM is present along that particular line-of-sight.

\section{\label{sec:concl}Conclusions}

\begin{figure}
\plotone{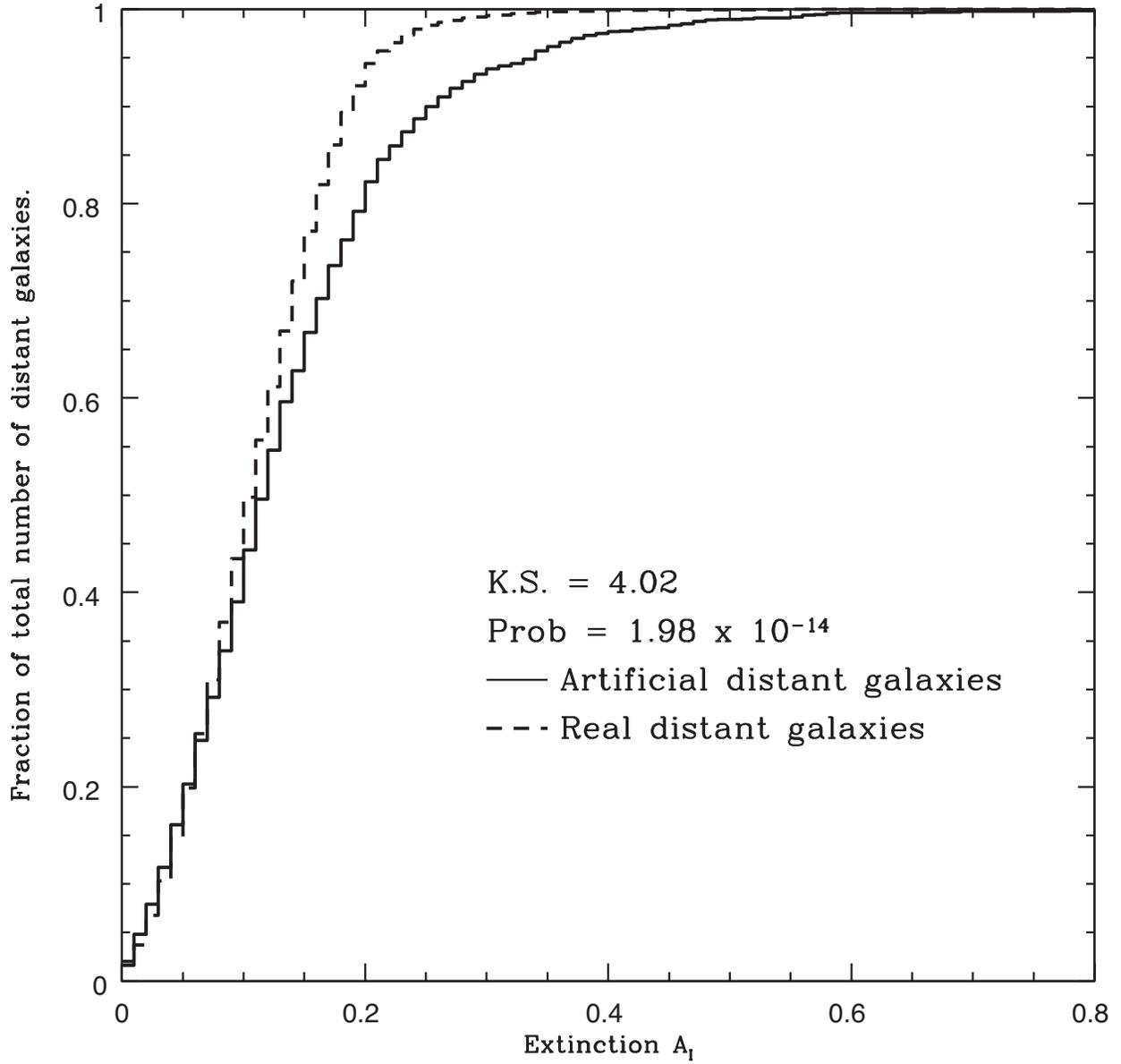}
\caption{\label{fig:hist}The cumulative histogram of extinction values at the position of observed 
distant background galaxies (dotted line histogram), compared to the histogram of extinction 
values based on the positions of artificial galaxies within the WFPC2 field-of-view (solid line histogram). }
\end{figure}

From the extinction values at the position of distant galaxies we draw the following conclusions:

\begin{itemize}
\item[1.] Distant galaxies are identified through a spiral disk in the lower-opacity 
regions. Their number is, only in part, determined by the cloud cover 
of large, resolved clouds in the disk (Figure \ref{fig:hist}). 
\item[2.] Most of the distant galaxies are identified in parts of the disk with some extinction. 
Dusty ISM clouds, unresolved by {\it Spitzer}, are the second cause diminishing their 
number (Figure \ref{fig:hist}).
\item[3.] The apparent optical depth from the number of distant galaxies can be 
expressed as a cloud cover fraction; however, this implies opaque clouds, while the 
majority of the dust clouds are optically thin and unresolved.
\item[4.] We reconfirm that extinction maps from an infrared color (e.g., E[I-L]) will mark 
the sections of higher extinction but underestimate the total extinction in any given part of the disk 
(Table \ref{tab:ext}) because of the intrinsic assumption of a simple geometry
\end{itemize}

In future applications of the counts of distant galaxies as a probe of disk 
opacity, the nature of the cloud geometry --expressed as a cloud 
covering factor-- should become increasingly evident.
We hope to learn more on the nature of the smooth extinction disk in 
our future studies using distant galaxies as an extinction test in the 
HST/ACS surveys of M51, M101, and M81.

\acknowledgements

The authors would like to thank T. Jarrett, for making his aperture corrections 
available, and B. Sugerman, for his help with the construction of the 
extinction maps.
The authors would like to thank the referee, D. Elmegreen, for her comments; 
they helped tremendously to improve the paper.
This work is based in part on archival data obtained with the {\it Spitzer} 
Space Telescope, which is operated by JPL, CalTech, under a contract 
with NASA. 
This work is also based on observations with the NASA/ESA Hubble Space 
Telescope, obtained at the STScI, which is operated by the Association 
of Universities for Research in Astronomy (AURA), Inc., under NASA 
contract NAS5-26555. 
Support for this work was provided by NASA through grant number 
c3886 to D. Calzetti. 


\begin{thebibliography}{86}
\expandafter\ifx\csname natexlab\endcsname\relax\def\natexlab#1{#1}\fi

\bibitem[{{Andredakis} \& {van der Kruit}(1992)}]{Andredakis92}
{Andredakis}, Y.~C. \& {van der Kruit}, P.~C. 1992, \aap, 265, 396

\bibitem[{{Baes} {et~al.}(2003){Baes}, {Davies}, {Dejonghe}, {Sabatini},
  {Roberts}, {Evans}, {Linder}, {Smith}, \& {de Blok}}]{Baes03}
{Baes}, M., {Davies}, J.~I., {Dejonghe}, H., {Sabatini}, S., {Roberts}, S.,
  {Evans}, R., {Linder}, S.~M., {Smith}, R.~M., \& {de Blok}, W.~J.~G. 2003,
  \mnras, 343, 1081

\bibitem[{{Baes} \& {Dejonghe}(2001{\natexlab{a}})}]{Baes01a}
{Baes}, M. \& {Dejonghe}, H. 2001{\natexlab{a}}, \mnras, 326, 722

\bibitem[{{Baes} \& {Dejonghe}(2001{\natexlab{b}})}]{Baes01b}
---. 2001{\natexlab{b}}, \mnras, 326, 733

\bibitem[{{Bell} \& {de Jong}(2001)}]{BelldeJong}
{Bell}, E.~F. \& {de Jong}, R.~S. 2001, \apj, 550, 212

\bibitem[{{Berlind} {et~al.}(1997){Berlind}, {Quillen}, {Pogge}, \&
  {Sellgren}}]{Berlind97}
{Berlind}, A.~A., {Quillen}, A.~C., {Pogge}, R.~W., \& {Sellgren}, K. 1997,
  \aj, 114, 107

\bibitem[{{Bianchi}(2007)}]{Bianchi07}
{Bianchi}, S. 2007, ArXiv e-prints, 705

\bibitem[{{Bianchi} {et~al.}(1996){Bianchi}, {Ferrara}, \&
  {Giovanardi}}]{Bianchi96}
{Bianchi}, S., {Ferrara}, A., \& {Giovanardi}, C. 1996, \apj, 465, 127

\bibitem[{{Boissier} {et~al.}(2004){Boissier}, {Boselli}, {Buat}, {Donas}, \&
  {Milliard}}]{Boissier04}
{Boissier}, S., {Boselli}, A., {Buat}, V., {Donas}, J., \& {Milliard}, B. 2004,
  \aap, 424, 465

\bibitem[{{Bruzual} {et~al.}(1988){Bruzual}, {Magris}, \& {Calvet}}]{Bruzual88}
{Bruzual}, A.~G., {Magris}, G., \& {Calvet}, N. 1988, \apj, 333, 673

\bibitem[{{Calzetti}(2001)}]{Calzetti01}
{Calzetti}, D. 2001, \pasp, 113, 1449

\bibitem[{{Calzetti} {et~al.}(2005){Calzetti}, {Kennicutt}, {Bianchi},
  {Thilker}, {Dale}, {Engelbracht}, {Leitherer}, {Meyer}, {Sosey}, {Mutchler},
  {Regan}, {Thornley}, {Armus}, {Bendo}, {Boissier}, {Boselli}, {Draine},
  {Gordon}, {Helou}, {Hollenbach}, {Kewley}, {Madore}, {Martin}, {Murphy},
  {Rieke}, {Rieke}, {Roussel}, {Sheth}, {Smith}, {Walter}, {White}, {Yi},
  {Scoville}, {Polletta}, \& {Lindler}}]{Calzetti05}
{Calzetti}, D., {Kennicutt}, R.~C., {Bianchi}, L., {Thilker}, D.~A., {Dale},
  D.~A., {Engelbracht}, C.~W., {Leitherer}, C., {Meyer}, M.~J., {Sosey}, M.~L.,
  {Mutchler}, M., {Regan}, M.~W., {Thornley}, M.~D., {Armus}, L., {Bendo},
  G.~J., {Boissier}, S., {Boselli}, A., {Draine}, B.~T., {Gordon}, K.~D.,
  {Helou}, G., {Hollenbach}, D.~J., {Kewley}, L., {Madore}, B.~F., {Martin},
  D.~C., {Murphy}, E.~J., {Rieke}, G.~H., {Rieke}, M.~J., {Roussel}, H.,
  {Sheth}, K., {Smith}, J.~D., {Walter}, F., {White}, B.~A., {Yi}, S.,
  {Scoville}, N.~Z., {Polletta}, M., \& {Lindler}, D. 2005, \apj, 633, 871

\bibitem[{{Calzetti} {et~al.}(1994){Calzetti}, {Kinney}, \&
  {Storchi-Bergmann}}]{Calzetti94}
{Calzetti}, D., {Kinney}, A.~L., \& {Storchi-Bergmann}, T. 1994, \apj, 429, 582

\bibitem[{{Cortese} {et~al.}(2005){Cortese}, {Boselli}, {Buat}, {Gavazzi},
  {Boissier}, {Gil de Paz}, {Seibert}, {Madore}, \& {Martin}}]{Cortese05}
{Cortese}, L., {Boselli}, A., {Buat}, V., {Gavazzi}, G., {Boissier}, S., {Gil
  de Paz}, A., {Seibert}, M., {Madore}, B.~F., \& {Martin}, D.~C. 2005, ArXiv
  Astrophysics e-prints

\bibitem[{{Dalcanton} {et~al.}(2004){Dalcanton}, {Yoachim}, \&
  {Bernstein}}]{Dalcanton04}
{Dalcanton}, J.~J., {Yoachim}, P., \& {Bernstein}, R.~A. 2004, \apj, 608, 189

\bibitem[{{Dale} {et~al.}(2005){Dale}, {Bendo}, {Engelbracht}, {Gordon},
  {Regan}, {Armus}, {Cannon}, {Calzetti}, {Draine}, {Helou}, {Joseph},
  {Kennicutt}, {Li}, {Murphy}, {Roussel}, {Walter}, {Hanson}, {Hollenbach},
  {Jarrett}, {Kewley}, {Lamanna}, {Leitherer}, {Meyer}, {Rieke}, {Rieke},
  {Sheth}, {Smith}, \& {Thornley}}]{Dale05}
{Dale}, D.~A., {Bendo}, G.~J., {Engelbracht}, C.~W., {Gordon}, K.~D., {Regan},
  M.~W., {Armus}, L., {Cannon}, J.~M., {Calzetti}, D., {Draine}, B.~T.,
  {Helou}, G., {Joseph}, R.~D., {Kennicutt}, R.~C., {Li}, A., {Murphy}, E.~J.,
  {Roussel}, H., {Walter}, F., {Hanson}, H.~M., {Hollenbach}, D.~J., {Jarrett},
  T.~H., {Kewley}, L.~J., {Lamanna}, C.~A., {Leitherer}, C., {Meyer}, M.~J.,
  {Rieke}, G.~H., {Rieke}, M.~J., {Sheth}, K., {Smith}, J.~D.~T., \&
  {Thornley}, M.~D. 2005, \apj, 633, 857

\bibitem[{{Dale} {et~al.}(2007){Dale}, {de Paz}, {Gordon}, {Hanson}, {Armus},
  {Bendo}, {Bianchi}, {Block}, {Boissier}, {Boselli}, {Buckalew}, {Buat},
  {Burgarella}, {Calzetti}, {Cannon}, {Engelbracht}, {Helou}, {Hollenbach},
  {Jarrett}, {Kennicutt}, {Leitherer}, {Li}, {Madore}, {Martin}, {Meyer},
  {Murphy}, {Regan}, {Roussel}, {Smith}, {Sosey}, {Thilker}, \&
  {Walter}}]{Dale07}
{Dale}, D.~A., {de Paz}, A.~G., {Gordon}, K.~D., {Hanson}, H.~M., {Armus}, L.,
  {Bendo}, G.~J., {Bianchi}, L., {Block}, M., {Boissier}, S., {Boselli}, A.,
  {Buckalew}, B.~A., {Buat}, V., {Burgarella}, D., {Calzetti}, D., {Cannon},
  J.~M., {Engelbracht}, C.~W., {Helou}, G., {Hollenbach}, D.~J., {Jarrett},
  T.~H., {Kennicutt}, R.~C., {Leitherer}, C., {Li}, A., {Madore}, B.~F.,
  {Martin}, D.~C., {Meyer}, M.~J., {Murphy}, E.~J., {Regan}, M.~W., {Roussel},
  H., {Smith}, J.~D.~T., {Sosey}, M.~L., {Thilker}, D.~A., \& {Walter}, F.
  2007, \apj, 655, 863

\bibitem[{{Dasyra} {et~al.}(2005){Dasyra}, {Xilouris}, {Misiriotis}, \&
  {Kylafis}}]{Dasyra05}
{Dasyra}, K.~M., {Xilouris}, E.~M., {Misiriotis}, A., \& {Kylafis}, N.~D. 2005,
  \aap, 437, 447

\bibitem[{{Davies} \& {Burstein}(1995)}]{Cardith94}
{Davies}, J.~I. \& {Burstein}, D., eds. 1995, {The opacity of spiral disks}

\bibitem[{{Disney} {et~al.}(1989){Disney}, {Davies}, \& {Phillipps}}]{Disney89}
{Disney}, M., {Davies}, J., \& {Phillipps}, S. 1989, \mnras, 239, 939

\bibitem[{{Domingue} {et~al.}(1999){Domingue}, {Keel}, {Ryder}, \&
  {White}}]{kw99a}
{Domingue}, D.~L., {Keel}, W.~C., {Ryder}, S.~D., \& {White}, R.~E. 1999, \aj,
  118, 1542

\bibitem[{{Domingue} {et~al.}(2000){Domingue}, {Keel}, \& {White}}]{kw00b}
{Domingue}, D.~L., {Keel}, W.~C., \& {White}, R.~E. 2000, \apj, 545, 171

\bibitem[{{Dopita} {et~al.}(2006{\natexlab{a}}){Dopita}, {Fischera},
  {Sutherland}, {Kewley}, {Leitherer}, {Tuffs}, {Popescu}, {van Breugel}, \&
  {Groves}}]{Dopita06b}
{Dopita}, M.~A., {Fischera}, J., {Sutherland}, R.~S., {Kewley}, L.~J.,
  {Leitherer}, C., {Tuffs}, R.~J., {Popescu}, C.~C., {van Breugel}, W., \&
  {Groves}, B.~A. 2006{\natexlab{a}}, \apjs, 167, 177

\bibitem[{{Dopita} {et~al.}(2006{\natexlab{b}}){Dopita}, {Fischera},
  {Sutherland}, {Kewley}, {Tuffs}, {Popescu}, {van Breugel}, {Groves}, \&
  {Leitherer}}]{Dopita06a}
{Dopita}, M.~A., {Fischera}, J., {Sutherland}, R.~S., {Kewley}, L.~J., {Tuffs},
  R.~J., {Popescu}, C.~C., {van Breugel}, W., {Groves}, B.~A., \& {Leitherer},
  C. 2006{\natexlab{b}}, \apj, 647, 244

\bibitem[{{Draine} {et~al.}(2007){Draine}, {Dale}, {Bendo}, {Gordon}, {Smith},
  {Armus}, {Engelbracht}, {Helou}, {Kennicutt}, {Li}, {Roussel}, {Walter},
  {Calzetti}, {Moustakas}, {Murphy}, {Rieke}, {Bot}, {Hollenbach}, {Sheth}, \&
  {Teplitz}}]{Draine07}
{Draine}, B.~T., {Dale}, D.~A., {Bendo}, G., {Gordon}, K.~D., {Smith},
  J.~D.~T., {Armus}, L., {Engelbracht}, C.~W., {Helou}, G., {Kennicutt}, Jr.,
  R.~C., {Li}, A., {Roussel}, H., {Walter}, F., {Calzetti}, D., {Moustakas},
  J., {Murphy}, E.~J., {Rieke}, G.~H., {Bot}, C., {Hollenbach}, D.~J., {Sheth},
  K., \& {Teplitz}, H.~I. 2007, \apj, 663, 866

\bibitem[{{Draine} \& {Li}(2001)}]{Draine01a}
{Draine}, B.~T. \& {Li}, A. 2001, \apj, 551, 807

\bibitem[{{Draine} \& {Li}(2006)}]{Draine06}
---. 2006, \apj

\bibitem[{{Elmegreen} \& {Block}(1999)}]{Elmegreen99}
{Elmegreen}, B.~G. \& {Block}, D.~L. 1999, \mnras, 303, 133

\bibitem[{{Elmegreen}(1980)}]{Elmegreen80}
{Elmegreen}, D.~M. 1980, \apjs, 43, 37

\bibitem[{{Elmegreen} {et~al.}(2001){Elmegreen}, {Kaufman}, {Elmegreen},
  {Brinks}, {Struck}, {Klari{\'c}}, \& {Thomasson}}]{Elmegreen01}
{Elmegreen}, D.~M., {Kaufman}, M., {Elmegreen}, B.~G., {Brinks}, E., {Struck},
  C., {Klari{\'c}}, M., \& {Thomasson}, M. 2001, \aj, 121, 182

\bibitem[{{Engelbracht} {et~al.}(2004){Engelbracht}, {Gordon}, {Bendo},
  {P{\'e}rez-Gonz{\'a}lez}, {Misselt}, {Rieke}, {Young}, {Hines}, {Kelly},
  {Stansberry}, {Papovich}, {Morrison}, {Egami}, {Su}, {Muzerolle}, {Dole},
  {Alonso-Herrero}, {Hinz}, {Smith}, {Latter}, {Noriega-Crespo}, {Padgett},
  {Rho}, {Frayer}, \& {Wachter}}]{Engelbracht04}
{Engelbracht}, C.~W., {Gordon}, K.~D., {Bendo}, G.~J.,
  {P{\'e}rez-Gonz{\'a}lez}, P.~G., {Misselt}, K.~A., {Rieke}, G.~H., {Young},
  E.~T., {Hines}, D.~C., {Kelly}, D.~M., {Stansberry}, J.~A., {Papovich}, C.,
  {Morrison}, J.~E., {Egami}, E., {Su}, K.~Y.~L., {Muzerolle}, J., {Dole}, H.,
  {Alonso-Herrero}, A., {Hinz}, J.~L., {Smith}, P.~S., {Latter}, W.~B.,
  {Noriega-Crespo}, A., {Padgett}, D.~L., {Rho}, J., {Frayer}, D.~T., \&
  {Wachter}, S. 2004, \apjs, 154, 248

\bibitem[{{Engelbracht} {et~al.}(2006){Engelbracht}, {Kundurthy}, {Gordon},
  {Rieke}, {Kennicutt}, {Smith}, {Regan}, {Makovoz}, {Sosey}, {Draine},
  {Helou}, {Armus}, {Calzetti}, {Meyer}, {Bendo}, {Walter}, {Hollenbach},
  {Cannon}, {Murphy}, {Dale}, {Buckalew}, \& {Sheth}}]{Engelbracht06}
{Engelbracht}, C.~W., {Kundurthy}, P., {Gordon}, K.~D., {Rieke}, G.~H.,
  {Kennicutt}, R.~C., {Smith}, J.-D.~T., {Regan}, M.~W., {Makovoz}, D.,
  {Sosey}, M., {Draine}, B.~T., {Helou}, G., {Armus}, L., {Calzetti}, D.,
  {Meyer}, M., {Bendo}, G.~J., {Walter}, F., {Hollenbach}, D., {Cannon}, J.~M.,
  {Murphy}, E.~J., {Dale}, D.~A., {Buckalew}, B.~A., \& {Sheth}, K. 2006,
  \apjl, 642, L127

\bibitem[{{Fischera} \& {Dopita}(2005)}]{Fischera05}
{Fischera}, J. \& {Dopita}, M. 2005, \apj, 619, 340

\bibitem[{{Fischera} {et~al.}(2003){Fischera}, {Dopita}, \&
  {Sutherland}}]{Fischera03}
{Fischera}, J., {Dopita}, M.~A., \& {Sutherland}, R.~S. 2003, \apjl, 599, L21

\bibitem[{{Gonz{\'a}lez} {et~al.}(1998){Gonz{\'a}lez}, {Allen}, {Dirsch},
  {Ferguson}, {Calzetti}, \& {Panagia}}]{Gonzalez98}
{Gonz{\'a}lez}, R.~A., {Allen}, R.~J., {Dirsch}, B., {Ferguson}, H.~C.,
  {Calzetti}, D., \& {Panagia}, N. 1998, \apj, 506, 152

\bibitem[{{Gonz{\'a}lez} {et~al.}(2003){Gonz{\'a}lez}, {Loinard}, {Allen}, \&
  {Muller}}]{Gonzalez03}
{Gonz{\'a}lez}, R.~A., {Loinard}, L., {Allen}, R.~J., \& {Muller}, S. 2003,
  \aj, 125, 1182

\bibitem[{{Gordon} {et~al.}(2006){Gordon}, {Bailin}, {Engelbracht}, {Rieke},
  {Misselt}, {Latter}, {Young}, {Ashby}, {Barmby}, {Gibson}, {Hines}, {Hinz},
  {Krause}, {Levine}, {Marleau}, {Noriega-Crespo}, {Stolovy}, {Thilker}, \&
  {Werner}}]{Gordon06}
{Gordon}, K.~D., {Bailin}, J., {Engelbracht}, C.~W., {Rieke}, G.~H., {Misselt},
  K.~A., {Latter}, W.~B., {Young}, E.~T., {Ashby}, M.~L.~N., {Barmby}, P.,
  {Gibson}, B.~K., {Hines}, D.~C., {Hinz}, J., {Krause}, O., {Levine}, D.~A.,
  {Marleau}, F.~R., {Noriega-Crespo}, A., {Stolovy}, S., {Thilker}, D.~A., \&
  {Werner}, M.~W. 2006, \apjl, 638, L87

\bibitem[{{Gordon} {et~al.}(2001){Gordon}, {Misselt}, {Witt}, \&
  {Clayton}}]{Gordon01}
{Gordon}, K.~D., {Misselt}, K.~A., {Witt}, A.~N., \& {Clayton}, G.~C. 2001,
  \apj, 551, 269

\bibitem[{{Gordon} {et~al.}(2004){Gordon}, {P{\'e}rez-Gonz{\'a}lez}, {Misselt},
  {Murphy}, {Bendo}, {Walter}, {Thornley}, {Kennicutt}, {Rieke}, {Engelbracht},
  {Smith}, {Alonso-Herrero}, {Appleton}, {Calzetti}, {Dale}, {Draine},
  {Frayer}, {Helou}, {Hinz}, {Hines}, {Kelly}, {Morrison}, {Muzerolle},
  {Regan}, {Stansberry}, {Stolovy}, {Storrie-Lombardi}, {Su}, \&
  {Young}}]{Gordon04}
{Gordon}, K.~D., {P{\'e}rez-Gonz{\'a}lez}, P.~G., {Misselt}, K.~A., {Murphy},
  E.~J., {Bendo}, G.~J., {Walter}, F., {Thornley}, M.~D., {Kennicutt}, Jr.,
  R.~C., {Rieke}, G.~H., {Engelbracht}, C.~W., {Smith}, J.-D.~T.,
  {Alonso-Herrero}, A., {Appleton}, P.~N., {Calzetti}, D., {Dale}, D.~A.,
  {Draine}, B.~T., {Frayer}, D.~T., {Helou}, G., {Hinz}, J.~L., {Hines}, D.~C.,
  {Kelly}, D.~M., {Morrison}, J.~E., {Muzerolle}, J., {Regan}, M.~W.,
  {Stansberry}, J.~A., {Stolovy}, S.~R., {Storrie-Lombardi}, L.~J., {Su},
  K.~Y.~L., \& {Young}, E.~T. 2004, \apjs, 154, 215

\bibitem[{{Holwerda}(2005)}]{Holwerda05}
{Holwerda}, B.~W. 2005, PhD thesis, Kapteyn Astronomical Institute

\bibitem[{{Holwerda} {et~al.}(2007){Holwerda}, {Draine}, {Gordon},
  {Gonz\'alez}, {Calzetti}, {Thornley}, B., {Allen}, \& {van der
  Kruit}}]{Holwerda07a}
{Holwerda}, B.~W., {Draine}, B., {Gordon}, K.~D., {Gonz\'alez}, R.~A.,
  {Calzetti}, D., {Thornley}, B., B., B., {Allen}, R.~J., \& {van der Kruit},
  P.~C. 2007, \aj, {\it submitted}

\bibitem[{{Holwerda} {et~al.}(2005{\natexlab{a}}){Holwerda}, {Gonz\'alez},
  {Allen}, \& {van der Kruit}}]{Holwerda05a}
{Holwerda}, B.~W., {Gonz\'alez}, R.~A., {Allen}, R.~J., \& {van der Kruit},
  P.~C. 2005{\natexlab{a}}, \aj, 129, 1381

\bibitem[{{Holwerda} {et~al.}(2005{\natexlab{b}}){Holwerda}, {Gonz\'alez},
  {Allen}, \& {van der Kruit}}]{Holwerda05b}
---. 2005{\natexlab{b}}, \aj, 129, 1396

\bibitem[{{Holwerda} {et~al.}(2005{\natexlab{c}}){Holwerda}, {Gonz\'alez},
  {Allen}, \& {van der Kruit}}]{Holwerda05c}
---. 2005{\natexlab{c}}, \aap, 444, 101

\bibitem[{{Holwerda} {et~al.}(2005{\natexlab{d}}){Holwerda}, {Gonz\'alez},
  {Allen}, \& {van der Kruit}}]{Holwerda05e}
---. 2005{\natexlab{d}}, \aap, 444, 319

\bibitem[{{Holwerda} {et~al.}(2006){Holwerda}, {Gonz\'alez}, {Calzetti},
  {Allen}, {van der Kruit}, \& {the SINGS team}}]{Holwerda06sings}
{Holwerda}, B.~W., {Gonz\'alez}, R.~A., {Calzetti}, D., {Allen}, R.~J., {van
  der Kruit}, P.~C., \& {the SINGS team}. 2006, ArXiv Astrophysics e-prints

\bibitem[{{Holwerda} {et~al.}(2005{\natexlab{e}}){Holwerda}, {Gonz\'alez}, {van
  der Kruit}, \& {Allen}}]{Holwerda05d}
{Holwerda}, B.~W., {Gonz\'alez}, R.~A., {van der Kruit}, P.~C., \& {Allen},
  R.~J. 2005{\natexlab{e}}, \aap, 444, 109

\bibitem[{{Howk}(1999)}]{Howk99a}
{Howk}, J.~C. 1999, \apss, 269, 293

\bibitem[{{Howk} \& {Savage}(1999)}]{Howk99b}
{Howk}, J.~C. \& {Savage}, B.~D. 1999, \aj, 117, 2077

\bibitem[{{Hunter} {et~al.}(2006){Hunter}, {Elmegreen}, \& {Martin}}]{Hunter06}
{Hunter}, D.~A., {Elmegreen}, B.~G., \& {Martin}, E. 2006, ArXiv Astrophysics
  e-prints

\bibitem[{{Jarrett}(2005)}]{IRACaper}
{Jarrett}, J. 2005, IRAC: Extended Source Calibration

\bibitem[{{Kamphuis} {et~al.}(2007){Kamphuis}, {Holwerda}, {Allen}, {Peletier},
  \& {van der Kruit}}]{Kamphuis07}
{Kamphuis}, P., {Holwerda}, B.~W., {Allen}, R.~J., {Peletier}, R.~F., \& {van
  der Kruit}, P.~C. 2007, \aa, 344, 868

\bibitem[{{Keel} \& {White}(2001{\natexlab{a}})}]{kw01a}
{Keel}, W.~C. \& {White}, R.~E. 2001{\natexlab{a}}, \aj, 121, 1442

\bibitem[{{Keel} \& {White}(2001{\natexlab{b}})}]{kw01b}
---. 2001{\natexlab{b}}, \aj, 122, 1369

\bibitem[{{Kennicutt} {et~al.}(2003){Kennicutt}, {Armus}, {Bendo}, {Calzetti},
  {Dale}, {Draine}, {Engelbracht}, {Gordon}, {Grauer}, {Helou}, {Hollenbach},
  {Jarrett}, {Kewley}, {Leitherer}, {Li}, {Malhotra}, {Regan}, {Rieke},
  {Rieke}, {Roussel}, {Smith}, {Thornley}, \& {Walter}}]{sings}
{Kennicutt}, R.~C., {Armus}, L., {Bendo}, G., {Calzetti}, D., {Dale}, D.~A.,
  {Draine}, B.~T., {Engelbracht}, C.~W., {Gordon}, K.~D., {Grauer}, A.~D.,
  {Helou}, G., {Hollenbach}, D.~J., {Jarrett}, T.~H., {Kewley}, L.~J.,
  {Leitherer}, C., {Li}, A., {Malhotra}, S., {Regan}, M.~W., {Rieke}, G.~H.,
  {Rieke}, M.~J., {Roussel}, H., {Smith}, J.~T., {Thornley}, M.~D., \&
  {Walter}, F. 2003, \pasp, 115, 928

\bibitem[{{Kodaira} \& {Ohta}(1994)}]{Kodaira94}
{Kodaira}, K. \& {Ohta}, K. 1994, \pasj, 46, 155

\bibitem[{{Kylafis} \& {Xilouris}(2005)}]{Kylafis05}
{Kylafis}, N.~D. \& {Xilouris}, E.~M. 2005, in American Institute of Physics
  Conference Series, Vol. 761, The Spectral Energy Distributions of Gas-Rich
  Galaxies: Confronting Models with Data, ed. C.~C. {Popescu} \& R.~J. {Tuffs},
  3--+

\bibitem[{{Li} \& {Draine}(2001)}]{Draine01b}
{Li}, A. \& {Draine}, B.~T. 2001, \apj, 554, 778

\bibitem[{{Lu} {et~al.}(2003){Lu}, {Helou}, {Werner}, {Dinerstein}, {Dale},
  {Silbermann}, {Malhotra}, {Beichman}, \& {Jarrett}}]{Lu03}
{Lu}, N., {Helou}, G., {Werner}, M.~W., {Dinerstein}, H.~L., {Dale}, D.~A.,
  {Silbermann}, N.~A., {Malhotra}, S., {Beichman}, C.~A., \& {Jarrett}, T.~H.
  2003, \apj, 588, 199

\bibitem[{{Meijerink} {et~al.}(2005){Meijerink}, {Tilanus}, {Dullemond},
  {Israel}, \& {van der Werf}}]{Meijerink05}
{Meijerink}, R., {Tilanus}, R.~P.~J., {Dullemond}, C.~P., {Israel}, F.~P., \&
  {van der Werf}, P.~P. 2005, \aap, 430, 427

\bibitem[{{Meixner} {et~al.}(2006){Meixner}, {Gordon}, {Indebetouw}, {Hora},
  {Whitney}, {Blum}, {Reach}, {Bernard}, {Meade}, {Babler}, {Engelbracht},
  {For}, {Misselt}, {Vijh}, {Leitherer}, {Cohen}, {Churchwell}, {Boulanger},
  {Frogel}, {Fukui}, {Gallagher}, {Gorjian}, {Harris}, {Kelly}, {Kawamura},
  {Kim}, {Latter}, {Madden}, {Markwick-Kemper}, {Mizuno}, {Mizuno}, {Mould},
  {Nota}, {Oey}, {Olsen}, {Onishi}, {Paladini}, {Panagia}, {Perez-Gonzalez},
  {Shibai}, {Sato}, {Smith}, {Staveley-Smith}, {Tielens}, {Ueta}, {Dyk},
  {Volk}, {Werner}, \& {Zaritsky}}]{sage}
{Meixner}, M., {Gordon}, K.~D., {Indebetouw}, R., {Hora}, J.~L., {Whitney}, B.,
  {Blum}, R., {Reach}, W., {Bernard}, J.-P., {Meade}, M., {Babler}, B.,
  {Engelbracht}, C.~W., {For}, B.-Q., {Misselt}, K., {Vijh}, U., {Leitherer},
  C., {Cohen}, M., {Churchwell}, E.~B., {Boulanger}, F., {Frogel}, J.~A.,
  {Fukui}, Y., {Gallagher}, J., {Gorjian}, V., {Harris}, J., {Kelly}, D.,
  {Kawamura}, A., {Kim}, S., {Latter}, W.~B., {Madden}, S., {Markwick-Kemper},
  C., {Mizuno}, A., {Mizuno}, N., {Mould}, J., {Nota}, A., {Oey}, M.~S.,
  {Olsen}, K., {Onishi}, T., {Paladini}, R., {Panagia}, N., {Perez-Gonzalez},
  P., {Shibai}, H., {Sato}, S., {Smith}, L., {Staveley-Smith}, L., {Tielens},
  A.~G.~G.~M., {Ueta}, T., {Dyk}, S.~V., {Volk}, K., {Werner}, M., \&
  {Zaritsky}, D. 2006, \aj, 132, 2268

\bibitem[{{Meyer} {et~al.}(2006){Meyer}, {Calzetti}, \& {Regan}}]{Martin06}
{Meyer}, M., {Calzetti}, D., \& {Regan}, M. 2006, \apj, 1, 1

\bibitem[{{Misiriotis} {et~al.}(2001){Misiriotis}, {Popescu}, {Tuffs}, \&
  {Kylafis}}]{Misiriotis01}
{Misiriotis}, A., {Popescu}, C.~C., {Tuffs}, R., \& {Kylafis}, N.~D. 2001,
  \aap, 372, 775

\bibitem[{{Misselt} {et~al.}(2001){Misselt}, {Gordon}, {Clayton}, \&
  {Wolff}}]{Misselt01}
{Misselt}, K.~A., {Gordon}, K.~D., {Clayton}, G.~C., \& {Wolff}, M.~J. 2001,
  \apj, 551, 277

\bibitem[{{Natta} \& {Panagia}(1984)}]{Natta84}
{Natta}, A. \& {Panagia}, N. 1984, \apj, 287, 228

\bibitem[{{Pahre} {et~al.}(2004{\natexlab{a}}){Pahre}, {Ashby}, {Fazio}, \&
  {Willner}}]{Pahre04a}
{Pahre}, M.~A., {Ashby}, M.~L.~N., {Fazio}, G.~G., \& {Willner}, S.~P.
  2004{\natexlab{a}}, \apjs, 154, 235

\bibitem[{{Pahre} {et~al.}(2004{\natexlab{b}}){Pahre}, {Ashby}, {Fazio}, \&
  {Willner}}]{Pahre04b}
---. 2004{\natexlab{b}}, \apjs, 154, 229

\bibitem[{{Peletier} {et~al.}(1995){Peletier}, {Valentijn}, {Moorwood},
  {Freudling}, {Knapen}, \& {Beckman}}]{Peletier95}
{Peletier}, R.~F., {Valentijn}, E.~A., {Moorwood}, A.~F.~M., {Freudling}, W.,
  {Knapen}, J.~H., \& {Beckman}, J.~E. 1995, \aap, 300, L1+

\bibitem[{{P\'{e}rez-Gonz\'{a}lez} {et~al.}(2006){P\'{e}rez-Gonz\'{a}lez},
  {Kennicutt}, {Gordon}, {Misselt}, {Gil de Paz}, {Engelbracht}, {Rieke},
  {Bendo}, {Bianchi}, {Boissier}, {Calzetti}, {Dale}, {Draine}, {Jarrett},
  {Hollenbach}, \& {Prescott}}]{PG06}
{P\'{e}rez-Gonz\'{a}lez}, P.~G., {Kennicutt}, Jr., R.~C., {Gordon}, K.~D.,
  {Misselt}, K.~A., {Gil de Paz}, A., {Engelbracht}, C.~W., {Rieke}, G.~H.,
  {Bendo}, G.~J., {Bianchi}, L., {Boissier}, S., {Calzetti}, D., {Dale}, D.~A.,
  {Draine}, B.~T., {Jarrett}, T.~H., {Hollenbach}, D., \& {Prescott}, M.~K.~M.
  2006, ArXiv Astrophysics e-prints

\bibitem[{{Popescu} {et~al.}(2000){Popescu}, {Misiriotis}, {Kylafis}, {Tuffs},
  \& {Fischera}}]{Popescu00}
{Popescu}, C.~C., {Misiriotis}, A., {Kylafis}, N.~D., {Tuffs}, R.~J., \&
  {Fischera}, J. 2000, \aap, 362, 138

\bibitem[{{Popescu} \& {Tuffs}(2005)}]{Popescu05rev}
{Popescu}, C.~C. \& {Tuffs}, R.~J. 2005, in AIP Conf. Proc. 761: The Spectral
  Energy Distributions of Gas-Rich Galaxies: Confronting Models with Data,
  155--+

\bibitem[{{Regan}(2000)}]{Regan00}
{Regan}, M.~W. 2000, \apj, 541, 142

\bibitem[{{Regan} {et~al.}(2004){Regan}, {Thornley}, {Bendo}, {Draine}, {Li},
  {Dale}, {Engelbracht}, {Kennicutt}, {Armus}, {Calzetti}, {Gordon}, {Helou},
  {Hollenbach}, {Jarrett}, {Kewley}, {Leitherer}, {Malhotra}, {Meyer},
  {Misselt}, {Morrison}, {Murphy}, {Muzerolle}, {Rieke}, {Rieke}, {Roussel},
  {Smith}, \& {Walter}}]{Regan04}
{Regan}, M.~W., {Thornley}, M.~D., {Bendo}, G.~J., {Draine}, B.~T., {Li}, A.,
  {Dale}, D.~A., {Engelbracht}, C.~W., {Kennicutt}, R.~C., {Armus}, L.,
  {Calzetti}, D., {Gordon}, K.~D., {Helou}, G., {Hollenbach}, D.~J., {Jarrett},
  T.~H., {Kewley}, L.~J., {Leitherer}, C., {Malhotra}, S., {Meyer}, M.,
  {Misselt}, K.~A., {Morrison}, J.~E., {Murphy}, E.~J., {Muzerolle}, J.,
  {Rieke}, G.~H., {Rieke}, M.~J., {Roussel}, H., {Smith}, J.-D.~T., \&
  {Walter}, F. 2004, \apjs, 154, 204

\bibitem[{{Rieke} \& {Lebofsky}(1985)}]{Rieke85}
{Rieke}, G.~H. \& {Lebofsky}, M.~J. 1985, \apj, 288, 618

\bibitem[{{Rix} \& {Rieke}(1993)}]{Rix93}
{Rix}, H. \& {Rieke}, M.~J. 1993, \apj, 418, 123

\bibitem[{{Seth} {et~al.}(2005){Seth}, {Dalcanton}, \& {de Jong}}]{Seth05}
{Seth}, A.~C., {Dalcanton}, J.~J., \& {de Jong}, R.~S. 2005, \aj, 130, 1574

\bibitem[{{SINGS team}(2006)}]{dr4}
{SINGS team}. 2006, SINGS fourth data delivery release notes, SINGS team

\bibitem[{{Thompson} {et~al.}(2004){Thompson}, {Howk}, \&
  {Savage}}]{Thompson04}
{Thompson}, T.~W.~J., {Howk}, J.~C., \& {Savage}, B.~D. 2004, \aj, 128, 662

\bibitem[{{Thornley} {et~al.}(2006){Thornley}, {Braine}, \&
  {Gardan}}]{Thornley06}
{Thornley}, M.~D., {Braine}, J., \& {Gardan}, E. 2006, \apjl, 651, L101

\bibitem[{{Tuffs} \& {Popescu}(2005)}]{Tuffs05rev}
{Tuffs}, R.~J. \& {Popescu}, C.~C. 2005, in AIP Conf. Proc. 761: The Spectral
  Energy Distributions of Gas-Rich Galaxies: Confronting Models with Data,
  344--+

\bibitem[{{Tuffs} {et~al.}(2004){Tuffs}, {Popescu}, {V{\"o}lk}, {Kylafis}, \&
  {Dopita}}]{Tuffs04}
{Tuffs}, R.~J., {Popescu}, C.~C., {V{\"o}lk}, H.~J., {Kylafis}, N.~D., \&
  {Dopita}, M.~A. 2004, \aap, 419, 821

\bibitem[{{White} \& {Keel}(1992)}]{kw92}
{White}, R.~E. \& {Keel}, W.~C. 1992, \nat, 359, 129

\bibitem[{{White} {et~al.}(2000){White}, {Keel}, \& {Conselice}}]{kw00a}
{White}, R.~E., {Keel}, W.~C., \& {Conselice}, C.~J. 2000, \apj, 542, 761

\bibitem[{{Witt} \& {Gordon}(2000)}]{Witt00}
{Witt}, A.~N. \& {Gordon}, K.~D. 2000, \apj, 528, 799

\bibitem[{{Witt} {et~al.}(1992){Witt}, {Thronson}, \& {Capuano}}]{Witt92}
{Witt}, A.~N., {Thronson}, H.~A., \& {Capuano}, J.~M. 1992, \apj, 393, 611

\bibitem[{{Xilouris} {et~al.}(1999){Xilouris}, {Byun}, {Kylafis}, {Paleologou},
  \& {Papamastorakis}}]{Xilouris99}
{Xilouris}, E.~M., {Byun}, Y.~I., {Kylafis}, N.~D., {Paleologou}, E.~V., \&
  {Papamastorakis}, J. 1999, \aap, 344, 868

\end{thebibliography}

\begin{deluxetable}{l l l}
\tablewidth{0pt.}
\tablecaption{ \label{tab:b} The zero-extinction [I-L] color; the zero-point $b$ in equation \ref{eq:extlaw}. The values display a range not unreasonable for spiral disk stellar populations \citep{BelldeJong}.}
\tablehead{ \colhead{Galaxy} 	& \colhead{E[I-L]}}
\startdata
NGC 0925	& -0.5 \\ 
NGC 2841	& -0.3 \\
NGC 3198	& -0.4 \\
NGC 3351	& -1.2 \\
NGC 3621	& -0.9 \\
NGC 3627	& -0.4 \\
NGC 4321	& -0.3 \\
NGC 4536	& -0.3 \\
NGC 4559	& -0.4 \\
NGC 4725	& -0.4\\
NGC 5194	& -0.6 \\
NGC 6946	& -0.3 \\
NGC 7331	& -0.3 \\
\enddata
\end{deluxetable}

\begin{deluxetable}{l l l l}
\tablewidth{0pt.}
\tablecaption{ \label{tab:ext} Average apparent extinction from the number of distant galaxies $A_{SFM}$, the Spitzer SED, $A_{SED}$, from \cite{Holwerda07a}; and the color maps in this work, $A_{[I-L]}$. These values have not been corrected for inclination, because such correction depends on the assumed dust geometry and the dust's effective filling factor.}
\tablehead{ \colhead{Galaxy} 	& \colhead{$A_{SFM}$}	& \colhead{$A_{SED}$} & \colhead{$A_{[I-L]}$}}
\startdata
NGC 0925		& $-0.4^{+0.3}_{-0.3}$	& 0.7	& 0.11 $\pm$ 0.07  \\
NGC 2841		& $0.8^{+0.5}_{-0.4}$	& 1.5	& 0.08 $\pm$ 0.05  \\
NGC 3198		& $0.8^{+0.3}_{-0.3}$	& 0.8	& 0.11 $\pm$ 0.07  \\
NGC 3351		& $ 1.2^{0.6}_{0.5}$		& 1.1	& \dots \\
NGC 3621-1		& $2.2^{+0.6}_{-0.6}$	& 1.1	& 0.08 $\pm$ 0.08  \\
NGC 3621-2		& $1.0^{+0.4}_{-0.3}$	& 0.8	& 0.09 $\pm$ 0.10  \\
NGC 3627		& $2.1^{+0.7}_{-0.7}$	& 1.8	& 0.12 $\pm$ 0.11  \\
NGC 4321		& $2.3^{+0.8}_{-0.7}$	& 3.0	& 0.06 $\pm$ 0.06  \\
NGC 4536		& $0.9^{+0.4}_{-0.4}$	& 0.3	& 0.11 $\pm$ 0.07  \\
NGC 4559		& $0.1^{+0.3}_{-0.3}$	& 0.3	& 0.08 $\pm$ 0.06  \\
NGC 4725		& $0.8^{+0.3}_{-0.3}$	& 0.5	& 0.05 $\pm$ 0.05  \\
NGC 5194-1		& $-0.4^{+0.4}_{-0.4}$	& 3.6	& 0.24 $\pm$ 0.08  \\
NGC 5194-2		& $1.4^{+0.6}_{-0.6}$	& 4.2	& 0.18 $\pm$ 0.09  \\
NGC 6946		& $1.1^{+0.6}_{-0.5}$	& 1.4	& 0.18 $\pm$ 0.11  \\
NGC 7331		& $0.3^{+0.3}_{-0.3}$	& 0.5	& 0.11 $\pm$ 0.08  \\
\enddata
\end{deluxetable}

\begin{deluxetable}{l l l l l l l l}
\tablewidth{0pt.}
\tablecaption{ \label{tab:f} Cloud covering percentages of the disk for different values of disk opacity ($\tau ~>$ 0.5, 0.3 and 0.1), both for the entire extinction map ($f_{\rm map}$) as well as for the section covered by the WFPC2 ($f_{\rm wfpc2}$). The last column is the cloud cover for the WFPC2 section implied from the SFM measurement, $f_{\rm SFM}$, assuming only optically thick clouds.}
\tablehead{ \colhead{Galaxy} 	& & \colhead{$f_{\rm map}$} & & & \colhead{$f_{\rm wfpc}$} & & \colhead{$f_{\rm SFM}$} \\
& \colhead{$>0.5$} & \colhead{$>0.3$} & \colhead{$>0.1$} & \colhead{$>0.5$} & \colhead{$>0.3$} & \colhead{$>0.1$} & 
}
\startdata
NGC 0925	& 0.2 & 1.6 & 38	& 0.1	& 4.9 	& 58	& \dots \\
NGC 2841	& 0.0 & 0.1 & 28	& 0.0	& 0.4		& 47	& 52\\
NGC 3198	& 0.1 & 1.3 & 55	& 0.3	& 3.3		& 74	& 52\\
NGC 3351	& 0.1 & 1.0 & 66	& 0.2	& 2.3		& 91	& 67\\
NGC 3621-1	& 1.0 & 5.1 & 46	& 0.3	& 2.8		& 40	& 87\\
NGC 3621-2	& 1.0 & 5.1 & 46	& 1.2	& 7.2		& 41	& 60\\
NGC 3627	& 0.5 & 3.4 & 39	& 2.4	& 8.2		& 72	& 86\\
NGC 4321	& 0.0 & 0.8 & 17	& 0.1	& 1.7		& 27	& 88\\
NGC 4536	& 0.7 & 2.7 & 17	& 0.0	& 3.9		& 61	& 56\\
NGC 4559	& 0.1 & 1.1 & 32	& 0.3	& 1.6		& 41	& 9\\
NGC 4725	& 0.0 & 0.1 & 10	& 0.0	& 0.5		& 22	& 52\\
NGC 5194-1	& 1.0 & 15 & 82		& 3.6	& 50		& 10	& \dots \\
NGC 5194-2	& 1.0 & 15 & 82		& 1.7	& 19		& 89	& 72\\
NGC 6946	& 1.0 & 11 & 87		& 4.0	& 22		& 94	& 64\\
NGC 7331	& 0.4 & 10 & 61		& 0.4	& 7.5		& 64	& 24\\
\enddata
\end{deluxetable}

\end{document}